\documentclass[twocolumn]{aastex62}
\usepackage{enumitem}
\usepackage[T1]{fontenc}
\usepackage{epsfig}
\usepackage{epstopdf}
\usepackage{natbib}
\usepackage{amssymb}
\usepackage{amsbsy}
\usepackage{natbib}
\usepackage{subfigure}
\usepackage[mathcal]{euscript}
\usepackage{float}
\usepackage{amsmath}
\usepackage{tabularx}
\usepackage{xspace}
\usepackage{enumitem}
\usepackage{color}
\usepackage{comment}
\usepackage{stackengine}

\usepackage{ulem,xspace}

\newcommand{\kms}{km~s$^{-1}$}
\newcommand{\vsini}{\ensuremath{v \sin{i}}}

\newcommand{\kepler}{{\it Kepler}}

\newcommand{\msun}{$\mathrm{M_{\odot}}$}

\newcommand{\lsun}{$\mathrm{L_{\odot}}$}

\newcommand{\mj}{$\mathrm{M_{\textrm{\scriptsize Jup}}}$}
\newcommand{\ms}{m~s$^{-1}$}
\newcommand{\gaia}{\textit{Gaia}}

\newcommand{\teff}{\mbox{$T_{\rm eff}$}}
\newcommand{\logg}{\mbox{$\log g$}}

\newcommand{\thekic}{{9267654}}

\bibpunct{(}{)}{;}{a}{}{,}

\begin{document}
\title{Spinning up the Surface: Evidence for Planetary Engulfment or Unexpected Angular Momentum Transport?}

\author[0000-0002-4818-7885]{Jamie Tayar}
\affiliation{Department of Astronomy, University of Florida, Bryant Space Science Center, Stadium Road, Gainesville, FL 32611, USA }
\affiliation{Institute for Astronomy, University of Hawai‘i at Mānoa, 2680 Woodlawn Drive, Honolulu, HI 96822, USA}
\author[0000-0002-9633-4093]{Facundo D. Moyano}
\affiliation{Observatoire de Gen\`eve, Universit\'e de Gen\`eve, 51 Ch. Pegasi, CH-1290 Versoix, Suisse}
\author[0000-0001-7493-7419]{Melinda Soares-Furtado}
\altaffiliation{NASA Hubble Fellow}
\affiliation{Department of Astronomy,  University of Wisconsin-Madison, 475 N. Charter St., Madison, WI 53703, USA}

\author[0000-0003-3833-2513]{Ana Escorza}
\affiliation{European Southern Observatory, Alonso de C\'{o}rdova 3107, Vitacura, Santiago, Chile}

\author[0000-0002-8717-127X]{Meridith Joyce}
\affiliation{Space Telescope Science Institute, 3700 San Martin Drive, Baltimore, MD, USA 21218}

\author[0000-0002-3430-4163]{Sarah~L.~Martell}
\affiliation{School of Physics, University of New South Wales, Sydney, NSW 2052, Australia}
\affiliation{Centre of Excellence for All-Sky Astrophysics in Three Dimensions (ASTRO 3D), Australia}

\author[0000-0002-8854-3776]{Rafael A. Garc\'\i a}
\affiliation{Université Paris-Saclay, Universit\'e Paris Cit\'e, CEA, CNRS, Astrophysique, Instrumentation et Modélisation Paris-Saclay, 91191 Gif-sur-Yvette Cedex, France}
\author[0000-0003-0377-0740]{Sylvain N. Breton}
\affiliation{Université Paris-Saclay, Universit\'e Paris Cit\'e, CEA, CNRS, Astrophysique, Instrumentation et Modélisation Paris-Saclay, 91191 Gif-sur-Yvette Cedex, France}
\author[0000-0001-9491-8012]{St\'ephane Mathis}
\affiliation{Université Paris-Saclay, Universit\'e Paris Cit\'e, CEA, CNRS, Astrophysique, Instrumentation et Modélisation Paris-Saclay, 91191 Gif-sur-Yvette Cedex, France}

\author[0000-0002-0129-0316]{Savita Mathur}
\affil{Instituto de Astrof\'isica de Canarias (IAC), E-38205 La Laguna, Tenerife, Spain}
\affil{Universidad de La Laguna (ULL), Departamento de Astrof\'isica, E-38206 La Laguna, Tenerife, Spain}

\author{Vincent Delsanti}
\affiliation{CentraleSup\'{e}lec, 3 Rue Joliot Curie, 91190 Gif-sur-Yvette, France}
\affiliation{AIM, CEA, CNRS, Universit\'e Paris-Saclay, Universit\'e de Paris, Sorbonne Paris Cit\'e, F-91191 Gif-sur-Yvette, France}

\author[0000-0003-1285-3433]{Sven Kiefer}
\affiliation{Institute of Astronomy, KU Leuven, Celestijnenlaan 200D, B-3001 Leuven, Belgium}
\affiliation{Space Research Institute, Austrian Academy of Sciences, Schmiedlstrasse 6, A-8042 Graz, Austria}
\affiliation{ TU Graz, Fakult\"at f\"ur Mathematik, Physik und Geod\"asie, Petersgasse 16, A-8010 Graz, Austria}

\author[0000-0002-0460-8289]{Sabine Reffert}
\affiliation{Landessternwarte, Zentrum für Astronomie der Universit\"at Heidelberg, K\"onigstuhl 12, 69117 Heidelberg, Germany}

\author[0000-0001-7402-3852]{Dominic M. Bowman}
\affiliation{Institute of Astronomy, KU Leuven, Celestijnenlaan 200D, B-3001 Leuven, Belgium}

\author[0000-0003-2771-1745]{Timothy Van Reeth}
\affiliation{Institute of Astronomy, KU Leuven, Celestijnenlaan 200D, B-3001 Leuven, Belgium}

\author[0000-0002-6244-0138]{Shreeya Shetye}
\affiliation{Institute of Physics, Laboratory of Astrophysics, École Polytechnique Fédérale de Lausanne (EPFL), Observatoire de Sauverny, 1290 Versoix, Switzerland}

\author{Charlotte Gehan}
\affiliation{Max-Planck-Institut f\"ur Sonnensystemforschung, Justus-von-Liebig-Weg 3, 37077 G\"ottingen, Germany}
\affiliation{Instituto de Astrof\'isica e Ci\^encias do Espaço, Universidade do Porto, CAUP, Rua das Estrelas, PT4150-762 Porto, Portugal}

\author[0000-0003-4976-9980]{Samuel K. Grunblatt}
\altaffiliation{Kalbfleisch Fellow}
\affiliation{American Museum of Natural History, 200 Central Park West, Manhattan, NY 10024, USA}
\affiliation{Center for Computational Astrophysics, Flatiron Institute, 162 5$^\text{th}$ Avenue, Manhattan, NY 10010, USA}

\begin{abstract}
In this paper, we report the potential detection of a nonmonotonic radial rotation profile in a low-mass lower-luminosity giant star. For most low- and intermediate-mass stars, the rotation on the main sequence seems to be close to rigid. As these stars evolve into giants, the core contracts and the envelope expands, which should suggest a radial rotation profile with a fast core and a slower envelope and surface. KIC \thekic, however, seems to show a surface rotation rate that is faster than its bulk envelope rotation rate, in conflict with this simple angular momentum conservation argument. We improve the spectroscopic surface constraint, show that the pulsation frequencies are consistent with the previously published core and envelope rotation rates, and demonstrate that the star does not show strong chemical peculiarities. We discuss the evidence against any tidally interacting stellar companion. Finally, we discuss the possible origin of this unusual rotation profile,  
including the potential ingestion of a giant planet or unusual angular momentum transport by tidal inertial waves triggered by a close substellar companion, and encourage further observational and theoretical efforts.  
\end{abstract}


\section{Introduction}
\setcounter{footnote}{0}
Stellar rotation rates are a sensitive tracer of stellar structure, magnetism, and internal transport of angular momentum. 
{As single stars evolve, several physical processes are expected to act, including angular momentum loss via stellar winds, meridional circulation, internal waves, and magnetic instabilities, among others \citep[see e.g.][]{MaederMeynet2000,Mathis2013LNP,Aerts2019}}. Unfortunately, their relative strength, effectiveness, and timescales have been challenging to constrain theoretically. Each mechanism is, however, generally expected to leave behind a characteristic radial rotation profile. 
{For example, in radiative regions purely hydrodynamic mechanisms} {such as shear instabilities} can act to reduce shear {while advective currents can create it} \citep[e.g.][]{Zahn1992,Talon1997,Meynet2000,Palacios2003,Decressin2009,Mathis2018}.
{Internal magnetic fields are often expected to lead to uniform rotation profiles in large stellar regions}  \citep[e.g.][]{Mestel1987,Spruit2002,Maeder2005,Eggenberger2005,Cantiello2014, Fuller2019}, while internal gravity waves can create complex and even counter rotating zones 
\citep{Kumar1999,Talon2005,Rogers2015}. 

The existence of large databases of rotation rates, including spot modulations \citep[e.g][]{ McQuillan2014, Ceillier2017,santos_2019, Gaulme2020, Santos2021} and spectroscopic \vsini\ measurements \citep[e.g][]{ZorecRoyer2012,Massarotti2008, Tayar2015} have significantly improved our theoretical predictions for the surface rotation rate evolution \citep{Matt2015, vanSaders2016, TayarPinsonneault2018}. In addition, the ability to use asteroseismology, the study of stellar oscillations, to constrain the internal rotation rates \citep{Beck2012, Mosser2012btest,Deheuvels2012, Deheuvels2014, VanReeth2016, Aerts2019, 2019LRSP...16....4G,Li2020,Pedersen2021} of these stars have greatly constrained the rates and timescales of angular momentum transport \citep{TayarPinsonneault2013, Cantiello2014} and led to significant theoretical work on the location of differential rotation and the mechanisms determining the rotation profile \citep{KissinThompson2015, Fuller2019, TakahashiLanger2021}. 

Stars enter the pre-main-sequence with a range of rotation rates inherited from their birth clouds. Interactions with their circumstellar disks as well as winds and outflows serve to regulate that angular momentum as they contract \citep[e.g.][]{CiezaBaliber2007}, and leave behind a distribution of rotation rates that is likely correlated with the properties of these disks \citep{Mamajek2009, Rebull2018}. At first, the core of these stars are decoupled from the angular momentum loss happening from the wind at the surface \citep{Denissenkov+2010}, but the two zones recouple on mass-dependent timescales of order tens to hundreds of millions of years as these stars approach the main sequence \citep{MacGregor1991, GalletBouvier2013, LanzafameSpada2015,SomersPinsonneault2016}. The complexity of these processes \citep[see also][]{Curtis2020, godoy_rivera_2021b} makes it challenging to predict \textit{a priori} what the rate of rotation or its radial or latitudinal profile should be for stars on the main sequence. 

Empirically, it has been found that stars rotate approximately as solid bodies on the main sequence \citep{VanReeth2016, VanReeth2018a, Li2019, Jermyn2020} and that seismic estimates of the envelope rotation rates in these stars are consistent with surface rotation measurements from spots or \vsini\ \citep{Nielsen2015, Benomar2015, Beck2018, hall2021}. 
Once stars leave the main sequence, they seem to rotate close to a solid body for a short evolutionary interval \citep{Deheuvels2020}, then the core contracts and the envelope expands, leading to a rapidly rotating core and a slowly rotating envelope.

However, it has been noted by many authors that the cores do not spin up to the level expected of angular momentum conservation \citep[see e.g.][]{TayarPinsonneault2013}, and so some mechanism must transport angular momentum from the core to the surface. Explorations of various hydrodynamic or magnetic mechanisms have thus far failed to match the necessary amount of angular momentum transport for all the observed stars \citep{Eggenberger2012, Ceillier2013, Marques2013, Cantiello2014}.

The identity of this mechanism has remained unclear, as no current theories perfectly match all the observations \citep[e.g.][]{Eggenberger2019,denharthog2020}. In addition, the radial location of the differential rotation is unclear, and various authors have argued that it could be in the surface convection zone \citep{KissinThompson2015, TayarPinsonneault2018, TakahashiLanger2021} although others expect it to be concentrated in radiative regions \citep{Fuller2019, Fellay2021}.
For most stars, getting measurements of the rotation rate at more than two locations- usually a `core' and a `surface'- has been challenging because of the limitations of the measurable oscillation modes, and thus this conflict has remained unresolved. 

In addition, it is known that orbital angular momentum can be exchanged with rotational angular momentum through tidal interactions \citep{Zahn1977}. 
Work has suggested that binary synchronization \citep{Song2013}, mass transfer \citep{Packet1981, RenzoGotberg2021}, and planet engulfment \citep{BolmontMathis2016,Privitera2016a, Privitera2016b,Ahuiretal2021} can all affect the rotation of a star.
However, the impact of these sorts of interactions on the stellar rotation profile, including the rate at which various regions of the star are spun up, is poorly constrained.

\section{KIC \thekic: Previous Analyses}

\begin{figure}[!htb]
\begin{center}
\def\stackalignment{l}
\subfigure{\topinset{\includegraphics[width=0.2\textwidth, clip=true, trim=0.5in 0in 0in 0in]{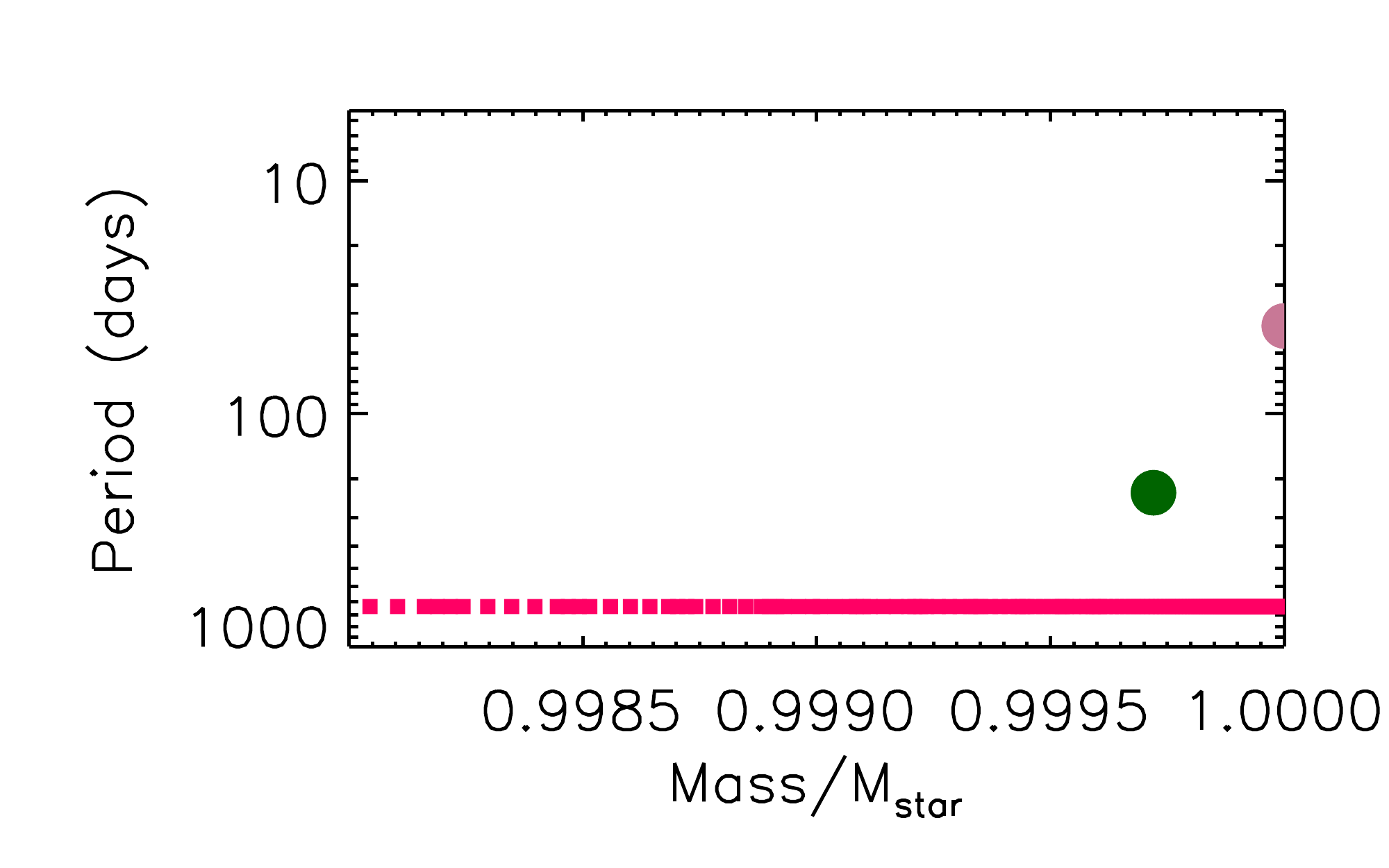}}{\includegraphics[width=8.5cm,clip=true, trim=0.5in 0in 0in 0in]{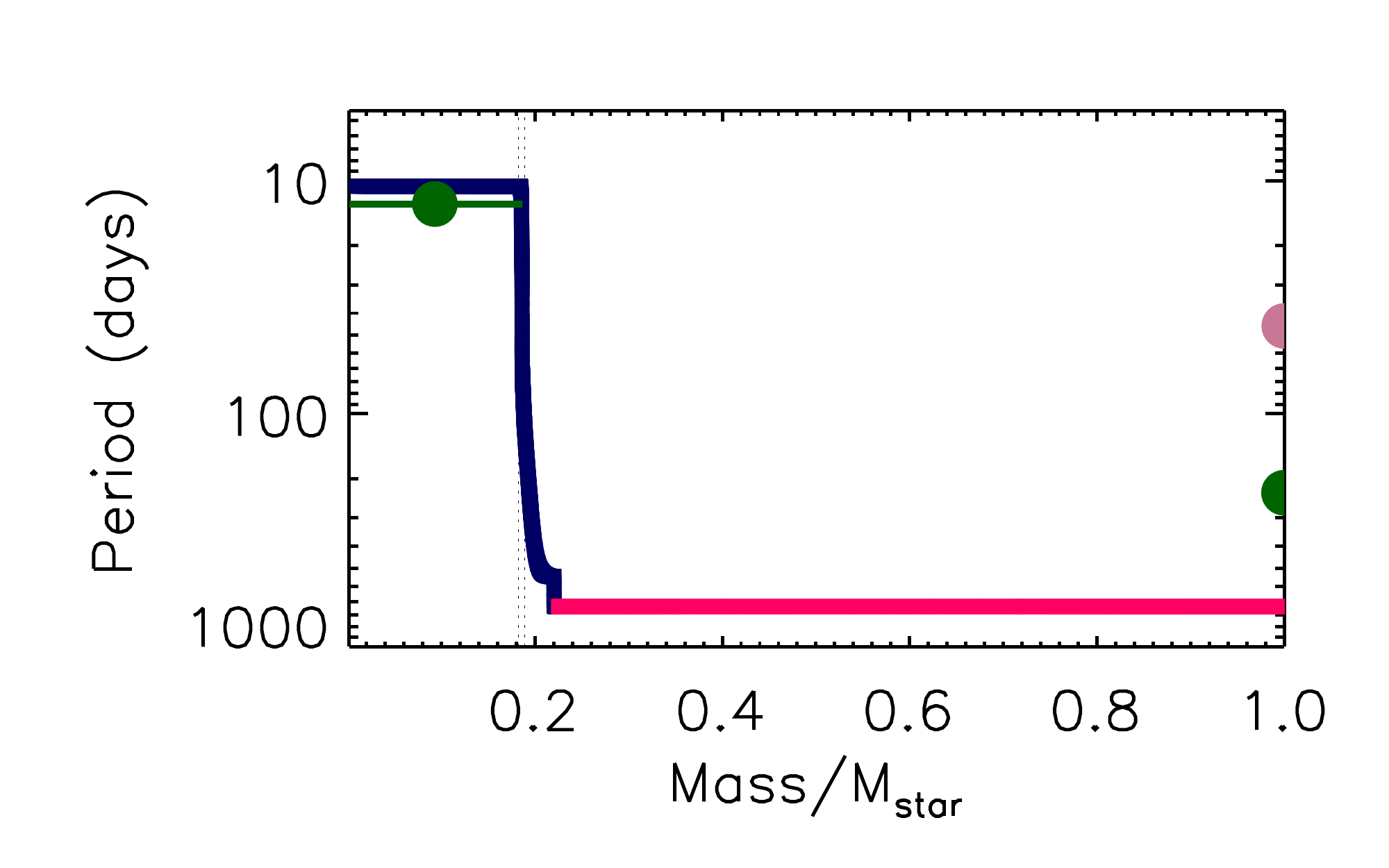}}{15pt}{110pt}}
\caption{Comparison of the inferred rotation rates as a function of the stellar mass coordinate (green points from asteroseismology, rose point from spectroscopy) to one theoretical model of the radial rotation profile \citep{Fuller2019}. We note that the model is included to provide guidance on what a plausible internal rotation profile might look like, but that any such model includes assumptions (internal magnetic field, initial surface velocity, etc.) that have not been tuned for this star specifically. Radiative shells are marked in blue and convective shells are marked in pink; vertical dashed lines mark the location of the hydrogen burning shell. The seismic points are qualitatively consistent with the model predictions, but the spectroscopic surface velocity is deeply inconsistent.}
\label{Fig:profile}
\end{center}
\end{figure}

In this paper, we present observations of KIC \thekic\ that suggest it has a nonmonotonic radial rotation profile (see Figure \ref{Fig:profile}). Asteroseismology, spectroscopy, and photometry agree that the star is a lower red giant branch (RGB) star (\teff$=4792\pm75$ K, \logg$=2.98\pm0.007$) of $1.138\pm0.039\textrm{ (random) }\pm0.035\textrm{ (systematic)}$ \msun\ \citep{Pinsonneault2018}. 
It was observed by the \kepler\ mission \citep{Borucki2010} for its full duration. Its seismic oscillations were first characterized by \citet{Mosser2012btest} and detailed modeling of its oscillation frequencies and rotational splittings were carried out by \citet{Corsaro2015a}, \citet{PerezHernandez2016}, and \citet{ Triana2017}. In particular, these works noted a well constrained core rotation rate of $\Omega_g$=908-939 nHz (P$_{\rm core}\sim$ 12.6 days), slightly higher than average for stars in this regime \citep{Gehan2018}.
Several methods of estimation suggested an envelope rotation rate of $\Omega_p$=20.8-65.0 nHz (P$_{\rm env}\sim$ 178-556 days), with most closer to the faster end of that estimate, although a linear fit to the approximate trapping parameter suggested a potentially counter-rotating envelope \citep[see][for more discussion of the various methods used to estimate the envelope rotation and their relative precision]{Triana2017}. We note that these results are consistent with the ensemble of core rotation rates available for similar stars \citep{Gehan2018}, other estimates of the core rotation rate of this star \citep[$\Omega_g$=920$\pm$10.06 nHz;][]{Gehan2018}, and also simple theoretical models of the surface rotational evolution. The oscillation pattern was used to estimate an inclination angle of 72.4$^{+17.6}_{-6.0}$ degrees for this star \citep{Gehan2021}, its modes, including the $\ell=1$ modes, looked to have normal visibility, and there was no evidence of surface spots or activity \citep{Ceillier2017,Gaulme2020}. This star was observed for two sectors by the Transiting Exoplanet Survey Satellite \citep[TESS;][]{Ricker2015}, and designated TIC 164832220, but it is too faint for oscillation modes to be easily detected in those data \citep[see][]{Stello2022}.

Very recently, \citet{MazzolaDaher2021} estimated surface rotation velocities for a large number of stars with spectroscopic data from the Sloan Digital Sky Survey \citep{Blanton2017} Apache Point Galactic Evolution Experiment \citep[APOGEE;][]{Majewski2017}.
These stars have moderate resolution (R $\sim$ 22,000) spectra taken in the H-band on the du Pont  and Sloan Foundation \citep{Gunn2006} telescopes and reduced automatically using the APOGEE Stellar Parameters and Chemical Abundances Pipeline \citep[ASPCAP;][]{Nidever2015, Zamora2015, GarciaPerez2016, Holtzman2015, Holtzman2018, Jonsson2020} to obtain spectroscopic parameters.
For red giant stars, the \citet{MazzolaDaher2021} analysis ran an additional search for rotational broadening using cross-correlation methods to artificially broadened template spectra \citep{Tayar2015, Dixon2020}. 
In their analysis, they suggested a relatively rapid rotation velocity of 6.25 \kms\ for this star, much faster than the ($<1$ \kms) rotation velocity suggested by the seismic envelope rotation and inclination constraints. 
They also found no strong evidence of a close binary companion in the APOGEE radial velocity measurements 
($\Delta$RV$_{\rm max}$= 0.3 \kms, v$_{\rm scatter}$=0.18 \kms), 
consistent with previous work by \citet{PriceWhelan2020}. 

The APOGEE survey also provided chemical abundance information for this star (see Section \ref{sec:chem}), and characterized it as a relatively common evolved solar-like object with a metallicity [Fe/H]$=-0.02\pm0.03$\,dex and an alpha-element enhancement [$\alpha$/Fe]$=0.00\pm0.01$, suggesting that it should be comparable to the ensemble of well studied stars in the \kepler\ field.
This star has also been observed by \gaia. Identified by the Gaia DR3 ID 2107170676942345088, its photometry is consistent with being a red giant \citep[G=11.97 and BP-RP=1.193;][]{Gaia_edr3} relatively nearby \citep[parallax=0.8444 milliarcseconds; distance 1151 pc;][]{Bailer-Jones2021}.
It follows a disk-like orbit with a low eccentricity (0.18), and it is currently about one thin disk scale height above the galactic plane (z$_{\rm max}=$ 408 pc).  
The Renormalized Unit Weight Error for this star is 0.969, which indicates that it is well fit as a single point source. 

\begin{figure}
\includegraphics[width=0.5\textwidth]{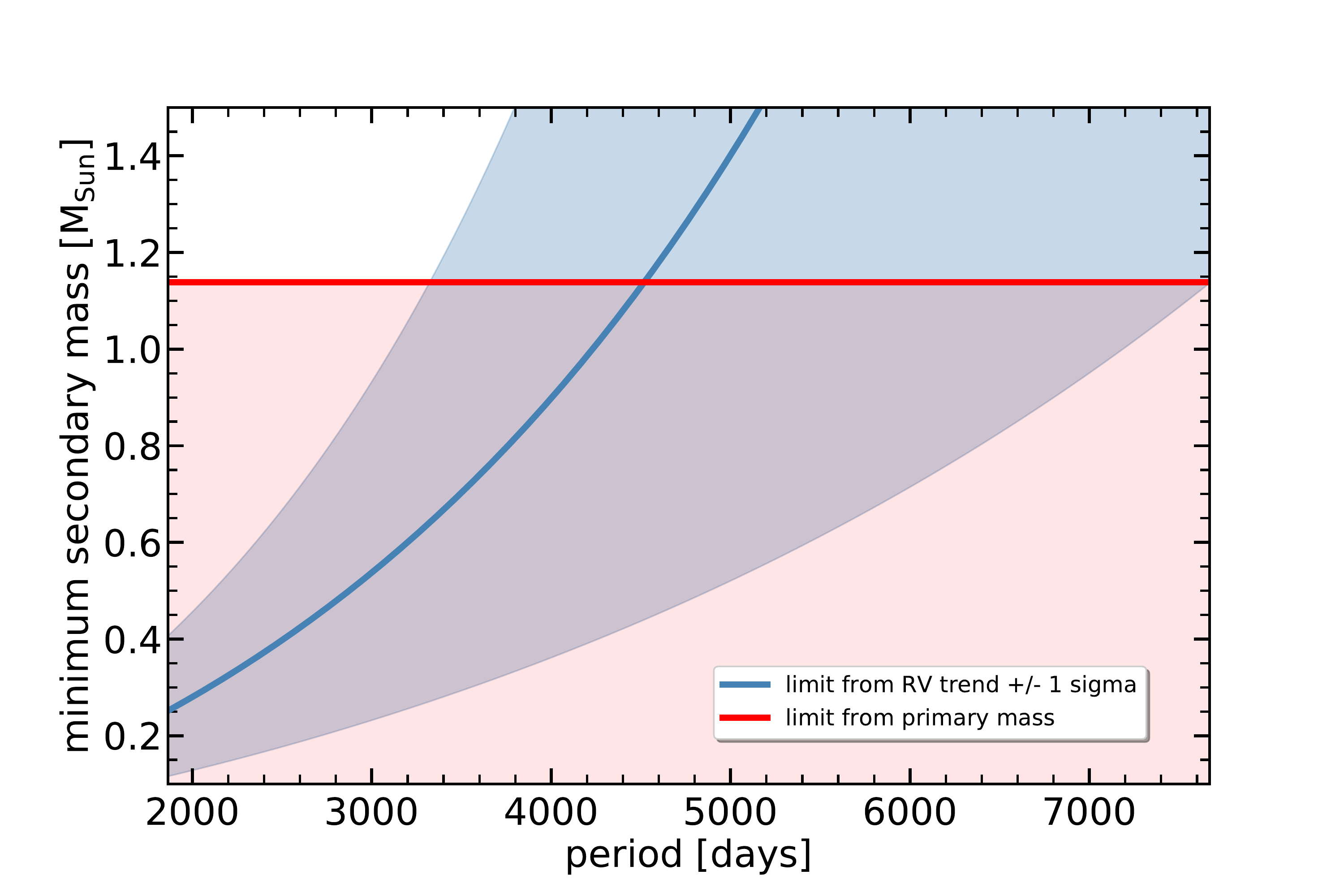}
\caption{Illustration of the periods and (minimum) secondary masses compatible with the radial velocity trend derived from \gaia\ DR3 radial velocities. The blue line indicates the minimum secondary mass derived from the radial velocity trend for each period, while the blue shaded region indicates the 1$\sigma$ confidence region. The red line denotes the primary mass, and the red shaded region the limits on the secondary mass based on the fact that the star is an SB1 only. The purple shaded overlap region indicates the combinations of period and secondary mass compatible with both constraints. Based on the nominal value for the radial velocity trend, periods roughly between five and twelve years and companion masses between 0.3 and 1.0~\msun\  are compatible with all constraints.}
\label{Fig:rvtrend}
\end{figure}

\section{Binary Analysis}

The \gaia\ radial velocity analysis published as part of the \gaia\ DR3 data release \citep{gaia_dr3} indicates a linear trend in the 24 radial velocities of KIC~\thekic\ of $-7.5 \pm 3.8$~\ms{}d$^{-1}$, significant at the 2$\sigma$ level. The star is denoted as an SB1, meaning there are no indications for a second set of lines and implying that the secondary must be fainter than the primary. It need not be much less massive than the primary though since the primary is evolved already, while the secondary might still be on the main-sequence. In the following, we derive limits on the mass of the secondary given the radial velocity trend and the primary mass. The 24 radial velocities measurements, which are not available yet individually, have been taken over a time interval of 931.915~days. The mean radial velocity of all measurements taken in that time frame is $-28.8\pm1.1$~\kms.

Assuming that a linear trend without notable deviations can be maintained over at most half of the period of an orbit, the minimum period of the system would be twice as long, i.e.\ 1864~days or about 5.1~years. For periods upwards of this limit, we have derived the constraint on the minimum companion mass (companion mass times the sine of the unknown orbital inclination) that would be compatible with a radial velocity semi-amplitude consistent with a linear trend of the given value over half of the assumed period under the assumption of a circular orbit. The blue line in Fig.~\ref{Fig:rvtrend} illustrates this constraint, while the blue area indicates its 1$\sigma$ confidence region. The red line indicates the primary mass; the secondary, assuming it is not only less massive but also less luminous than the primary, should have a somewhat smaller mass. As can be seen, orbital periods between 5.1~years and about 12~years are all compatible with the nominal observed radial velocity trend, and would generate minimum secondary masses between about 0.3 and 1.0\,\msun. If the error on the radial velocity trend is taken into account, the minimum masses could be even smaller (down to about 0.1\,\msun) and the orbital periods somewhat larger (up to about 21 years). Ground based radial velocity measurements from APOGEE and the Large Sky Area Multi-Object Fiber Spectroscopic Telescope \citep[LAMOST;][]{LAMOST} are sparse and uncertain enough that while they are consistent with the presence of such a trend at about the same order of magnitude, they do not add significant additional constraints. The minimum astrometric signature of all of these orbits is too small (at maximum of the order of 20~microarcseconds for the widest orbits) to be detected by \gaia, so this is consistent with the star not having been identified as an astrometric binary already. Depending on the inclination, the astrometric signature could in theory be larger than this limit, and the astrometric excess noise of 63~microarcseconds observed for this star could in principle be due to the spectroscopic companion. However, since the \gaia\ astrometric residuals also suffer from systematics, we refrain from modeling the orbit further based on the astrometric jitter. 

We conclude that it is highly likely that KIC \thekic\ is orbited by a low-mass stellar companion with a period of the order of 5 to 20 years. However, it is complicated to use such a companion to directly explain the surface rotation period. In general, tidal forces (see Section \ref{ssec:tides} for more discussion) work towards synchronizing the orbital and rotational motion, so in the case of a circular orbit, a companion with a 5 year period would be both too far away to be strongly interacting and would be forcing the surface towards a 5 year rotation period, not the 42 day period observed. In eccentric orbits, the rotation rate is spun up towards the pseudosynchronous rotation rate, although it does not always reach that value on short timescales \citep{Zimmerman2017}. In order for the likely companion inferred from \gaia\ to be pseudosynchronized with the observed surface rotation would require an orbital eccentricity greater that 0.98. While highly eccentric binaries including giants do exist \citep[e.g.][]{Heeren2021}, the system properties require careful tuning to survive the main sequence phase, so we work under the assumption that the wide companion is not directly responsible for the observed rotation. There are also orbital configurations that lead to more complex dynamics where a wide companion can tighten the orbit of an inner pair \citep[e.g. the eccentric Kozai-Lidov effect;][]{Naoz2016}, and it is the unidentified inner body that is causing the anomalous surface rotation. Such a scenario is not inconsistent with our observations, but we do not have sufficient data to confirm the presence of any inner companion directly.

We therefore conclude that the companion identified in the \gaia\ data should be well separated from and significantly fainter than the star of interest, so it should not substantially impact the detailed seismic or spectroscopic analyses discussed in the following sections. We also do not expect it to be directly related to the unusual rotation profile we observe, but suggest that further study could prove interesting in a dynamical sense.


\begin{figure}[!thb]
\begin{center}
\subfigure{\includegraphics[width=3.5in,clip=true, trim=0.5in 0.3in 5.5in 3.75in]{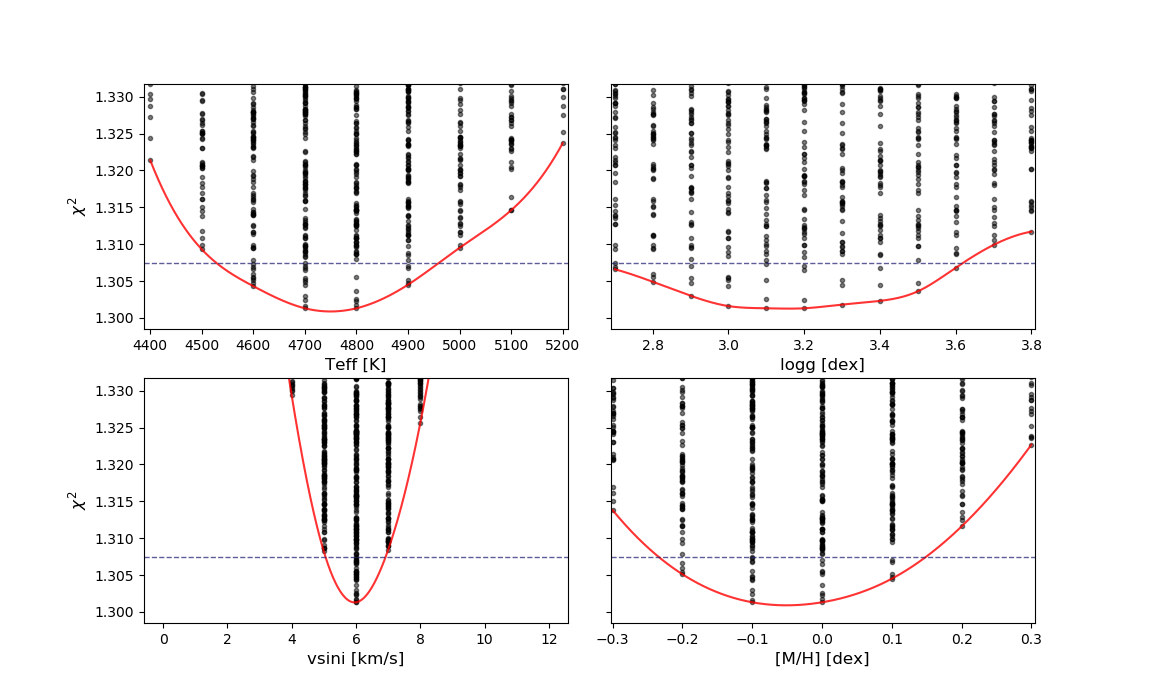}}
\subfigure{\includegraphics[width=3.5in,clip=true, trim=0.4in 0.2in 0.6in 0.8in]{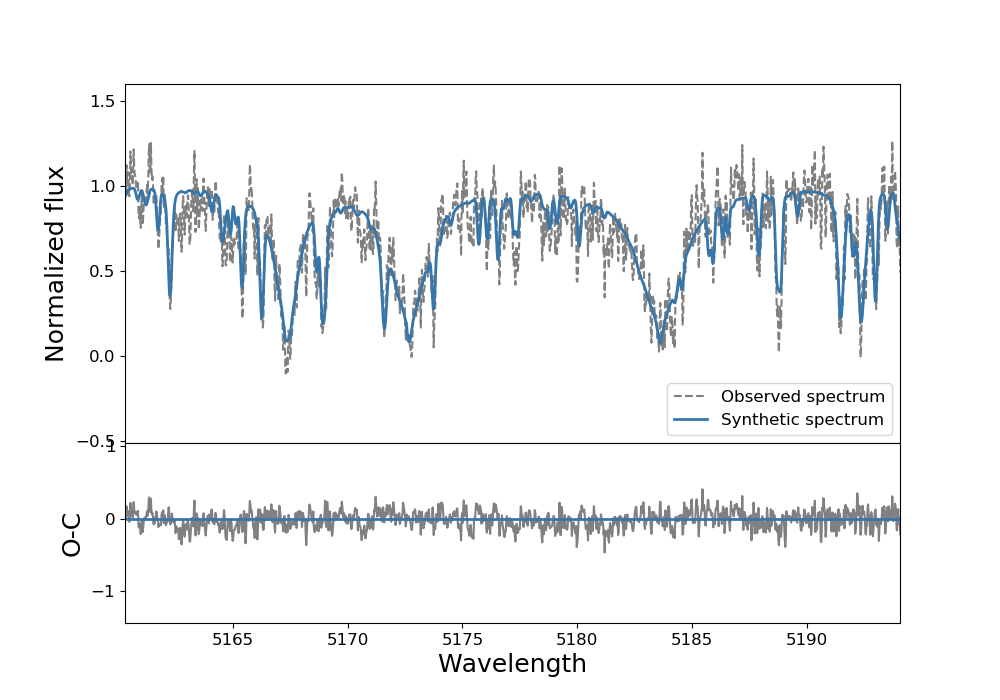}}

\subfigure{\includegraphics[width=3.5in,clip=true, trim=0.4in 0.2in 0.6in 0.8in]{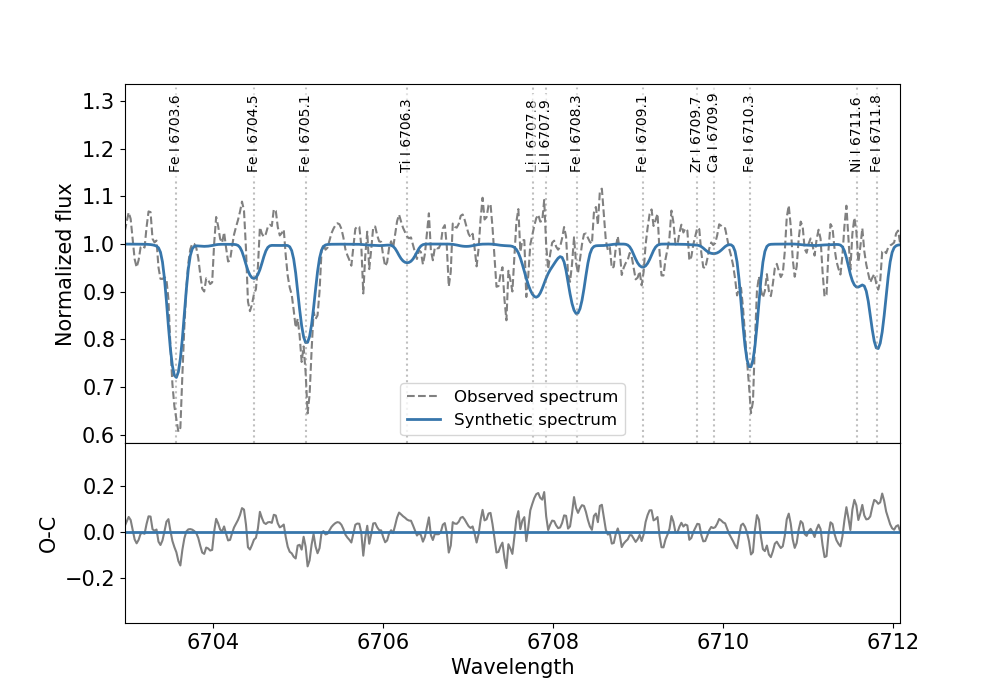}}
\caption{\textbf{Top:} $\chi$-squared minimization of the fit to \vsini\ in the HERMES spectrum gives a value of 5.96$^{+0.91}_{-0.97}$ \kms, with points indicating combinations of temperature, metallicity, gravity, and \vsini\ that were tested, and lower values of $\chi$-squared indicating better fits.
\textbf{Middle:}  Comparison to the magnesium triplet region suggests stellar parameters that are well fit and consistent with the published APOGEE results.
\textbf{Bottom:}  We also note that the lithium 6707 feature is not detected in this spectrum, indicating that the lithium abundance is subsolar. }
\label{Fig:spectra}
\end{center}
\end{figure}

\section{Spectroscopic Analysis}

Given the highly unusual detection of a surface rotation faster than the bulk envelope rate, it seemed necessary to confirm the surface rotation velocity. In particular, the APOGEE resolution is only R $\sim$ 22000, which has a formal \vsini\ detection limit of 13.1 \kms. While cross-correlation analysis has suggested that it is possible to distinguish slower rotation velocities, values below 10 \kms\ are still difficult, and most authors are highly skeptical of constraints below $\sim$ 5-8 \kms\ \citep{Tayar2015, Simonian2020, Dixon2020}. 

We therefore obtained a higher resolution spectrum using the High-Efficiency and high-Resolution Mercator Echelle Spectrograph \citep[HERMES;][]{Raskin2011} instrument on the 1.2-m Mercator telescope on La Palma. The star was observed on October 28th, 2021 for two observations of 45 min, each with a signal-to-noise ratio of $\approx$20 at 618\,nm and with a resolution of R$\simeq$ 85\,000. Data were reduced using the dedicated HERMES reduction pipeline, which includes flat-fielding and wavelength calibration \citep{Raskin2011}. In order to infer the spectroscopic parameters, the HERMES spectra were first normalized by fitting a second order spline to the continuum following the method described in \cite{Abdul-Masih2021}. Then the two spectra were co-added and compared to a large grid of model atmospheres using the spectral analysis code Grid Search in Stellar Parameters \citep[GSSP;][]{Tkachenko2015}. GSSP uses a grid of atmosphere models pre-computed with \textsc{LLmodels} \citep{LLmodels} and computes synthetic spectra on the fly using the \textsc{SynthV} \citep{Tsymbal1996} spectrum synthesis code. The HERMES spectral range covers from 400 to 900\,nm, but only the region from 450 to 700 was used for the stellar parameter search. This analysis was done independently of the APOGEE parameters.

The resulting fits (see Figure \ref{Fig:spectra}) were very similar to the APOGEE values with \teff $= 4750^{+220}_{-210}$ K, \logg = 3.16 $\pm$ 0.46 dex, and [M/H] $= -0.05^{+0.18}_{-0.20}$ dex, suggesting that the published spectroscopic parameters are likely to be reliable. In addition, this analysis  
gave an estimate for the surface rotation velocity of \vsini $= 5.96^{+0.91}_{-0.97}$ \kms, 
consistent with the APOGEE constraint, and significantly inconsistent with the seismic envelope rotation rate.

In addition, the cross-correlation function of the individual HERMES spectra show no evidence for a second set of lines that would indicate a spectroscopic binary companion, nor do they show any strong radial velocity shifts that may have been expected in a tidally locked system with a stellar mass secondary component, although we cannot confirm or further constrain the sort of wide companion suggested by the \gaia\ data. We also report the nondetection of lithium features at around 670\,nm in our optical spectrum. Given the measurement constraints, this rules out a lithium abundance at or above Solar. Such abundances would be consistent with previous observations of the lithium abundance for a star of this mass, metallicity, and surface gravity.  However, given that we only measure an upper limit, we cannot rule out this star being depleted in lithium, which has been suggested to occur in some cases where planetary engulfment drives mixing through the thermohaline instability \citep{Deal2015}.

\section{Asteroseismic Analysis}


\begin{figure}[!tb]
\begin{center}
\subfigure{\includegraphics[width=9cm,clip=true, trim=0in 0.0in 0in 0.in]{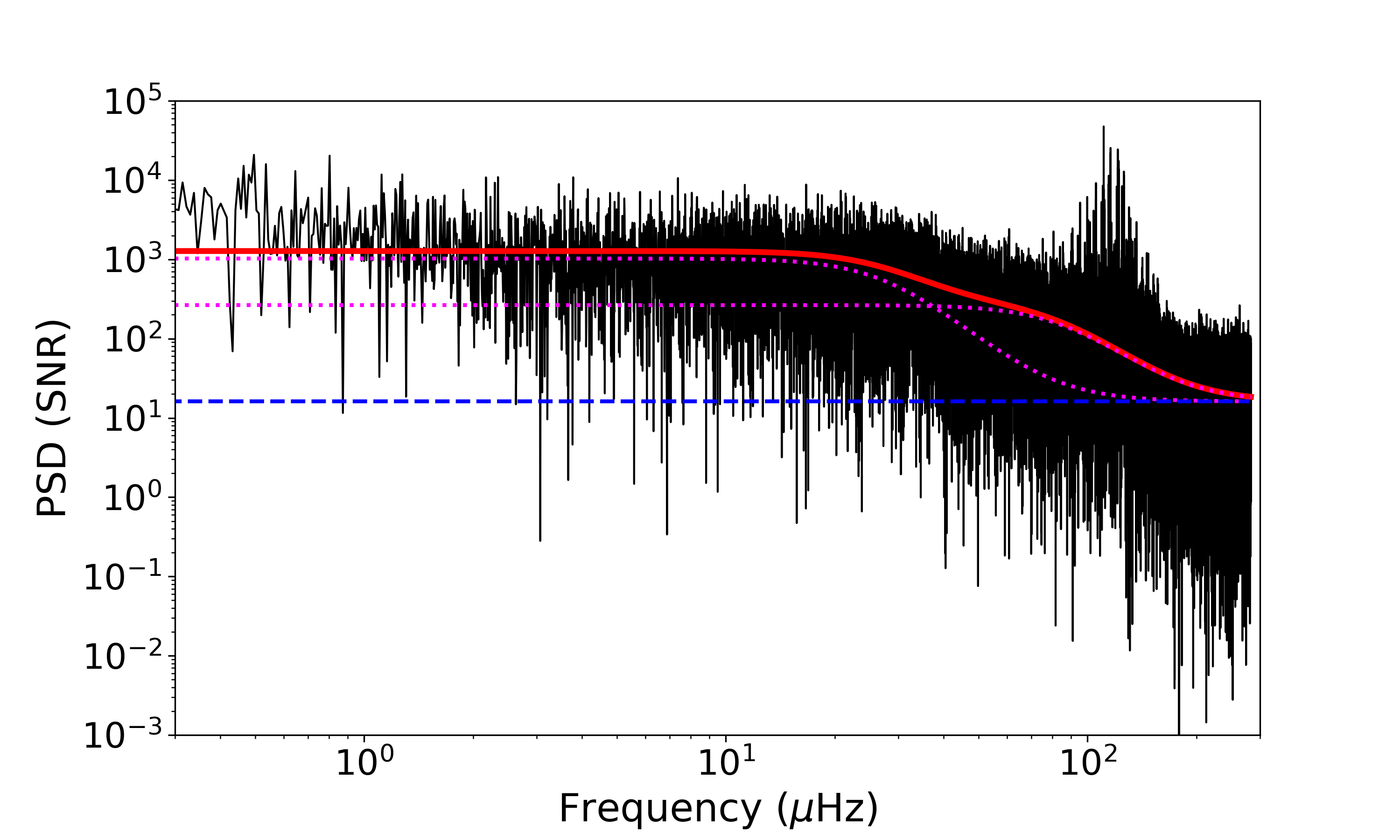}}
\subfigure{\includegraphics[width=9cm,clip=true, trim=0in 0in 0.0in 0.2in]{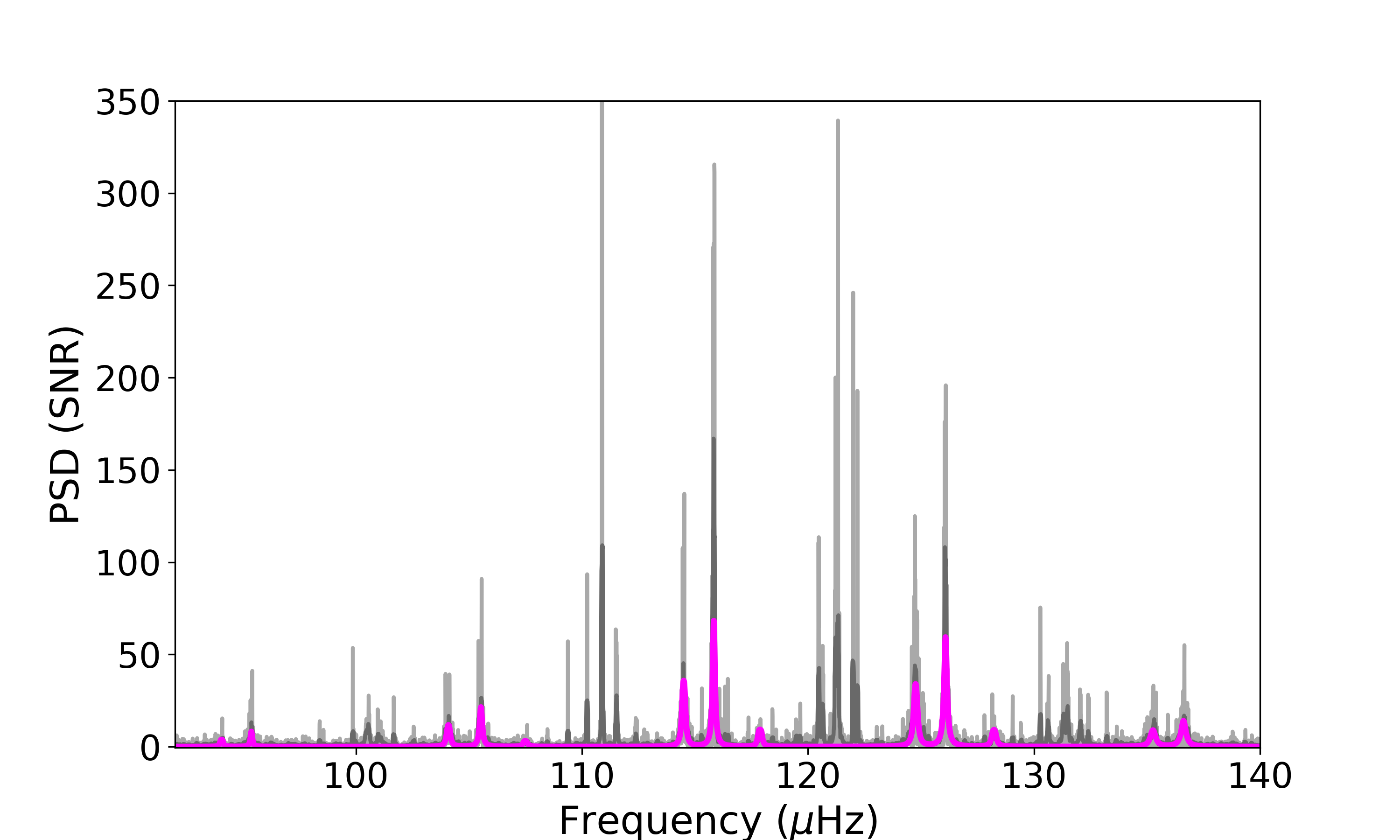}} 

\caption{\textbf{Top:} Power Spectrum Density (PSD) of KIC~9267654 (black) in logarithmic scale. Red line represent the sum of the fitted background components. Magenta dotted lines are the two convective contributions. The white noise is the blue dashed line. \textbf{Bottom:} Zoom of the above Power Spectrum Density in linear scale around $\nu_{\rm{max}}$ (in light grey), and a smoothed version by an eleven-point boxcar function in dark gray. Both PSDs are plotted in SNR after dividing the PSD by the background model (red continuous line on the top plot). The magenta line is the fitted model containing modes with degrees $\ell$=0, 2, and 3. Mixed dipolar modes were not fitted. 
}

\label{Fig:seismo}
\end{center}
\end{figure}

The original asteroseismic estimates of the core and envelope rotation rates come from the work of \citet{Triana2017}, which analyzed a sample of 13 stars on the lower red giant branch using a variety of methods to try to separate the information from the core region from the envelope region. Their core rotation estimate was later shown to be consistent with an entirely independent analysis published in \citet{Gehan2018}, emphasizing that this result, which is much simpler to measure, is likely to be robust. However, the envelope rotation rate can be more challenging to estimate, and so they applied 6 different methods including models, inversions, inferences, and fits \citep[see][for more detailed discussion of each method]{Triana2017}. For KIC \thekic, four of the rotation rates measured by these methods agreed to within the uncertainties. 

For this paper, we have taken a careful look at the available \kepler\ data, using the KEPSEISMIC\footnote{\url{https://archive.stsci.edu/prepds/kepseismic/}} light curves \citep{Garcia2011, Garcia2014, Pires2015}, where the data systematics, gaps, and instrumental effects have been carefully treated, and compare the results to the ensemble of \kepler\ data that have been homogeneously analyzed in the time since the original papers on this star were published. In our analysis, we have searched for any indication that would suggest that any of the properties might be unusual or mismeasured in a way that could impact the conclusions of this paper. We have found no such evidence.

Specifically, we find that the convective background for this star (see top panel in Fig.~\ref{Fig:seismo})  is entirely consistent with the relations found for the \emph{Kepler} red giants \citep{Mathur2011, Kallinger2014}, in agreement with what was found for this star in \citet{Sayeed2021}. This also makes it hard to imagine that the spectroscopic broadening we are interpreting as rotation is caused by unusual atmospheric macroturbulence, which should also correlate with the stellar surface convection and thus the granulation. We also find that the mode amplitudes and widths for this star are fully consistent with the expectations for a star of this type \citep[e.g.][]{Mosser2011b, Corsaro2017b}, with no obvious indications of e.g. strong magnetic fields \citep{Fuller2015,Loi2020b, Bugnet2021}. All of these properties are measured at extremely high signal to noise (SNR, see bottom panel in Fig.~\ref{Fig:seismo}) using the Markov Chain Monte Carlo (MCMC) Bayesian fitting code \texttt{apollinaire}\footnote{\texttt{https://apollinaire.readthedocs.io/en/latest/}} \citep{Breton2022}.  By doing so, we ensured that the new results are not dependent on the  Bayesian fitting scheme used \citep[i.e. MCMC with \texttt{apollinaire} versus a nested sampling tool in DIAMONDS;][]{CorsaroDeRidder2014}.

Because the signal to noise in the seismic modes is so high, estimating rotational splittings for this star is much less challenging than usual. The g-mode dominated $\ell=1$ mixed modes are significantly narrower and offset from the usual $\ell=1$ ridge; their splitting is clear and consistent with previous estimates of the core rotation rate.

\begin{figure}[ht!]
    \centering
    \includegraphics[width=9cm,clip=true]{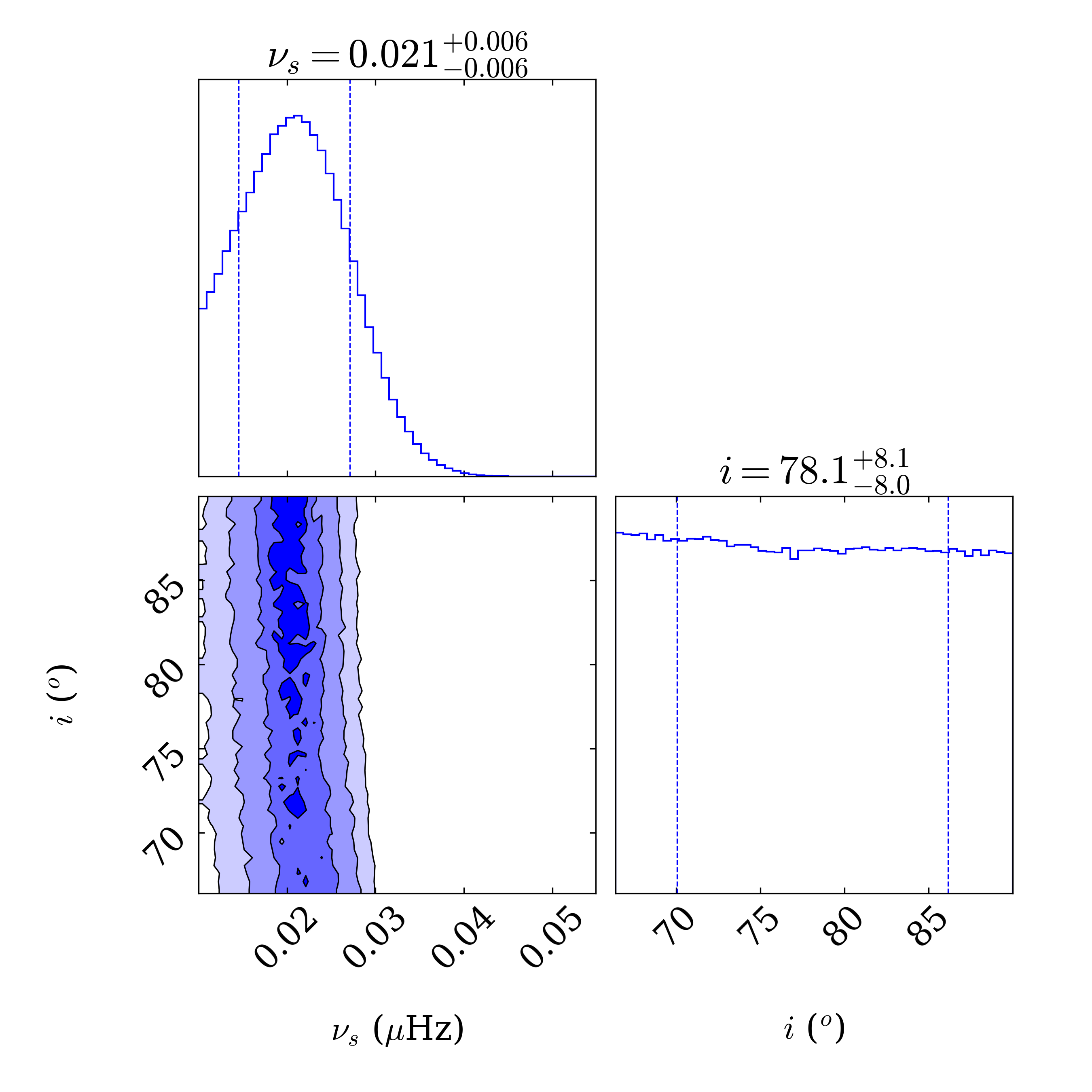}
    \caption{Marginalised posterior probability distribution obtained from the \texttt{apollinaire} MCMC sampling, for the mean rotational splitting $\nu_s$ (\textit{top}) and the inclination angle $i$ (\textit{bottom right}). The bottom left panel show the marginalised posterior probability on the couple $\nu_s$-$i$ and allow visualising the covariance of the two parameters.}
    \label{modep_split}
\end{figure}

We have also looked at the available $\ell=2$ and $\ell=3$ modes for potential detection of rotational splittings from the envelope region. Using \texttt{apollinaire}, {we fit 5 orders of $\ell=0,2$ modes and three $\ell=3$ modes between 95 and 136 $\mu$Hz. We allow the splitting, $\nu_s$, to vary between 0.01 and 0.6 $\mu$Hz, and allow the amplitude ratios between the different degree modes to vary freely.
Our MCMC chain had 500 walkers and between 20,000 and 55,000 steps. We tested a variety of different constraints on the inclination angle.
We concluded that it is not possible with the current data to effectively constrain the inclination angle of the star using only the p modes. It was only possible to constrain a combination of the projected splitting and the inclination angle \citep{BallotGarcia2006}. For this reason, we used the inclination angle derived from the mixed modes by \citet{Gehan2018} to run a final fit of the p modes; the result is shown in Fig.~\ref{modep_split}. This requires assuming that the inclination of the envelope is the same as the inclination of the core, which may or may not be true depending on what has happened to this star. Nevertheless, this seems the most reasonable starting assumption, and it produces a posterior probability of the inclination angle is rather flat with a median value of $78.1^{+8.1}_{-8.0}$ degrees and a rather Gaussian posterior probability for the rotational splitting centered at 0.021$\pm$0.006 $\mu$Hz (corresponding to rotation periods between 445 to 825 days). It is worth noting that for all of the fits using only p modes that we have tried, the envelope rotation periods obtained were longer than 200 days, regardless of the chosen inclination constraint. They were also generally slower than the rotation constraints published in \citet{Triana2017}, but for our purposes that only makes the detected rapid rotation of the surface \textit{more} significant. }

Further explorations of this star might consider imposing a nonmonotonic rotation profile \textit{ad hoc} on top of a stellar structure model whose fundamental parameters reproduce those of KIC 9267654. The imposed rotation profile should match the inferred estimates at the core-envelope boundary, subsurface convection zone, and surface, respectively. From here, one could determine that model's theoretical pulsation spectrum using the GYRE  stellar oscillation program \citep{townsend2013}, including the rotation kernels of each oscillation mode observed in the spectrum,  and compare this to the observations  to determine whether additional constraints can be placed on the stellar latitudinal or radial rotation profile \citep{Garcia2008, Davies2015}. However, given the challenges of such an exercise \citep{Gizon2003}, we consider it
beyond the scope of this study.

\section{Chemical Offsets}\label{sec:chem}

It has been shown that 
much of the variance in the elemental abundances of dwarf stars in the Galactic disk is confined to a small number of independent chemical vectors 
that can be mapped to various galactic enrichment processes \citep{PriceJonesBovy2018,Weinberg2019, Griffith2021, Joyce2022}. This provides us with fairly strong expectations for the abundance patterns of stars as a function of age, mass, and metallicity, making it possible to identify outliers from the expected pattern in a fairly straightforward manner.
If the unusual rotation profile of KIC \thekic\ is due to some kind of mass transfer or collision with a companion or a planet, enrichment by an AGB star, or the result of formation in an unusual environment, it is conceivable that such an event could have resulted in a deviation of its surface abundances from this expected pattern. Again, we find no strong evidence of such a deviation. 

\begin{figure}[!htb]
\begin{center}
{\includegraphics[width=9cm]{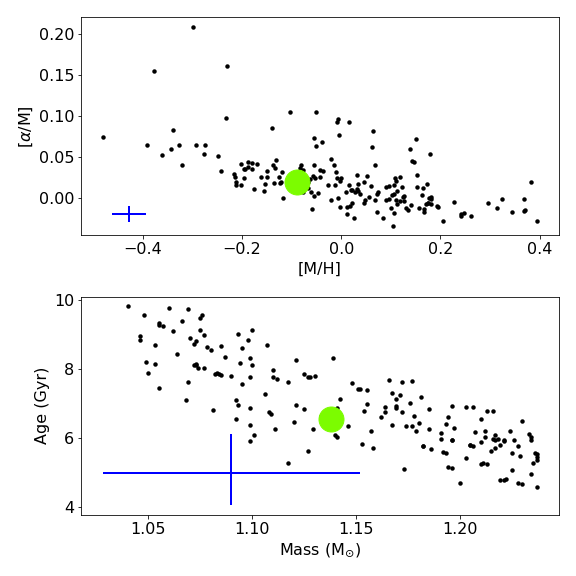}}
\caption{\textbf{Top:} Metallicity versus $\alpha$-enhancement for KIC \thekic\ (green circle) and the comparison stars from APOKASC (smaller black circles). \textbf{Bottom:} Mass versus age for KIC \thekic\ and the comparison stars from APOKASC. The blue cross in the lower left corner of each panel shows the median uncertainty for the comparison stars. The reported APOGEE uncertainties in the fundamental stellar properties [M/H] and [$\alpha$/M] are expanded by a factor of 5 for visibility.}
\label{Fig:comps_2}
\end{center}
\end{figure}

To explore this possibility, we selected a comparison sample from the APOKASC-2 catalogue of stars with measured asteroseismic gravities, ages, and masses \citep{Pinsonneault2018} and many elemental abundances determined in APOGEE DR16 \citep{DR16}. We kept all stars with valid results for seismic \logg, age, and mass, and with \teff{} within 50~K, \logg\, within 0.3 dex, and mass within $0.1~M_{\odot}$ of KIC \thekic. Figure  \ref{Fig:comps_2} compares these 176 RGB stars to KIC \thekic\ in $\alpha$-enhancement versus metallicity and mass versus age, to give a sense of the properties of this star as well as the range of the comparison sample, and they are all quite similar in these fundamental quantities. As one might expect when looking at a restricted metallicity range in the \kepler\ field, most of these stars are $\alpha$-poor disk members with $\alpha$-element abundances that correlate slightly with metallicity, and masses that are strongly correlated with their inferred ages. 

\begin{figure}[!htb]
\begin{center}
{\includegraphics[width=9cm]{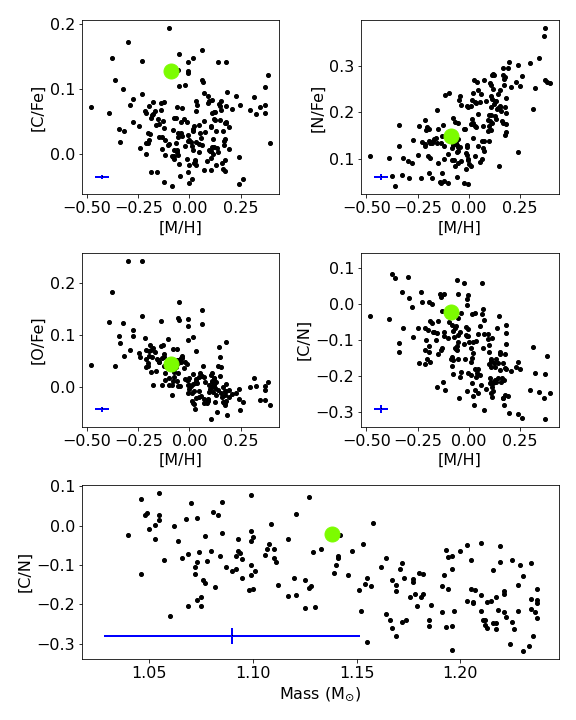}}
\caption{\textbf{Top Panels}: Abundance versus metallicity for the light elements C, N and O in KIC \thekic\ (green circle) and the comparison stars from APOKASC (smaller black circles). \textbf{Bottom:} The carbon-to-nitrogen ratio correlates well with stellar mass in the sample. The comparison of KIC \thekic\ to the trend suggests it could not have gained more than a few hundredths of a solar mass since its first dredge up. The blue cross in the lower left corner of each panel shows the median uncertainty for the comparison stars. As in Figure \ref{Fig:comps_2}, the uncertainty in [M/H] is expanded by a factor of 5 for visibility.}
\label{Fig:comps_4}
\end{center}
\end{figure}

\begin{figure*}[!htb]
\begin{center}
{\includegraphics[width=16cm]{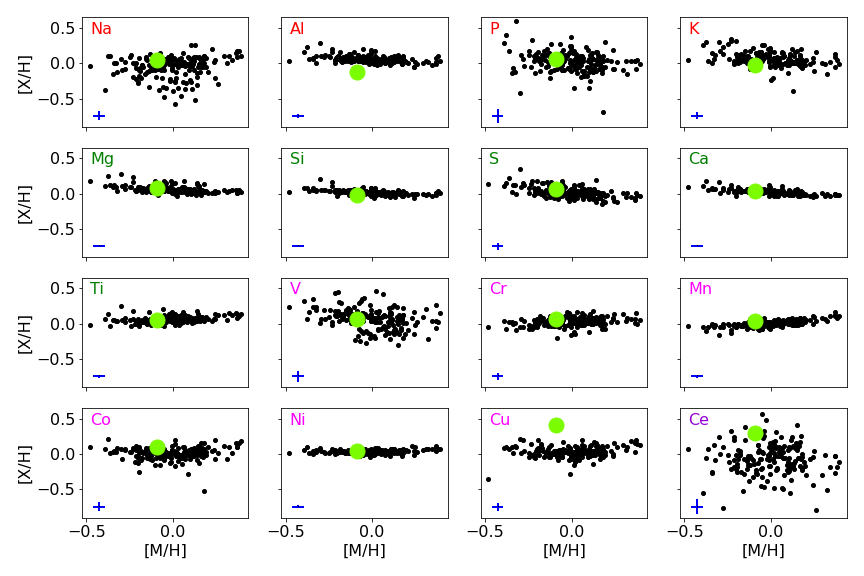}}
\caption{Abundance versus metallicity for the remaining elements from APOGEE in KIC \thekic\ (green circle) and the comparison stars from APOKASC (smaller black circles). In each panel, the element is identified in the upper left corner, with light odd-Z elements labeled in red, alpha elements labeled in green, iron peak elements labeled in magenta, and neutron-capture elements labeled in purple. The abundance pattern of KIC \thekic\ is consistent with the comparison sample in all elements except for Al, where it is depleted, and Cu, where it is enhanced. The blue cross in the lower left corner of each panel shows the median uncertainty for the comparison stars. As in Figures \ref{Fig:comps_2} and \ref{Fig:comps_4}, the uncertainty in [M/H] is expanded by a factor of 5 for visibility.}
\label{Fig:comps_5}
\end{center}
\end{figure*}

In Figures \ref{Fig:comps_2}, \ref{Fig:comps_4} and \ref{Fig:comps_5} the blue crosses in the lower left of each panel show the median uncertainties in the comparison set, and we have increased the reported uncertainties in the fundamental parameters [M/H] and [$\alpha$/Fe] by a factor of 5 to make them more visible. While we note that the reported uncertainties seem rather small, investigating their reasonableness is outside the scope of the present investigation.

Figures \ref{Fig:comps_4} and \ref{Fig:comps_5} show individual elemental abundances versus metallicity for the comparison stars as well as KIC \thekic. Figure \ref{Fig:comps_4} focuses on carbon, nitrogen and oxygen, which are produced and destroyed in CNO cycle fusion, and serve as sensitive indicators of dredge-up and internal mixing in red giant stars. The value of the carbon-to-nitrogen ratio [C/N], which is set by the depth and temperature of the first dredge up in giants, is strongly correlated with stellar mass for stars low on the red giant branch, as this star is \citep{Martig2016, Ness2016}. The gain of a large amount of mass from a stellar or substellar companion could potentially alter the surface abundances and move a star off of this correlation. We  show in the lower panel of Fig. \ref{Fig:comps_4} that this star is only slightly discrepant with the comparison sample, and as such has likely not gained more than a few hundredths of a solar mass since its first dredge-up. The middle right panel in Fig. \ref{Fig:comps_4} also shows [C/N] versus [M/H]. The ratio for KIC \thekic\ is well within the distribution of the APOKASC comparison sample, which similarly
does not seem to require significant abundance changes or mass gain from a companion after the first dredge up. 

In Figure \ref{Fig:comps_5} elements from the different nucleosynthetic groups are labeled in red (light odd-Z), green (alpha), magenta (iron peak) and purple (neutron capture). KIC \thekic\ is clearly depleted in aluminium and enhanced in copper relative to the comparison set, and mildly enhanced in carbon, manganese and cerium and depleted in silicon and potassium. These are not known signals of any common internal mixing, binary mass transfer, or planetary engulfment process.

We also explore whether the lack of detected lithium enhancement can place constraints on the accretion of material from a stellar or substellar companion.
Using the GALactic Archaeology with HERMES DR2 dataset \citep[GALAH; ][]{GalahDR2}, merged with effective temperature and luminosity measurements from Gaia DR2, we investigated a population of 26,183 stars with lithium abundance signatures and $\mathrm{[Fe/H]}=0 \pm 1$~dex.
Following the procedure outlined in \cite{Soares-Furtado2021}, we binned these signatures by the stellar effective temperature and luminosity, calculating the median lithium baseline for each bin, as well as the median absolute deviation of the median lithium baseline.
Using a post-processing approach that is also outlined in \cite{Soares-Furtado2021}, we then determined the lithium enrichment signature accompanying the ingestion of a 1~\mj{} substellar companion. 

For a star of 1.1~\msun{},
$T_\mathrm{eff}=4792$~K, and $L=15.4$~\lsun{}, we calculated that a lithium baseline of $A(\mathrm{Li})=1.03\pm 0.18$~dex would be expected on average, although a large range of lithium abundances is observed in the population. 
For such a host, the ingestion of a 1~\mj{} companion would result in a lithium abundance signature of $1.1$~dex, which would only represent a statistically insignificant $0.5{\sigma}$ increase in the expected lithium abundance. 

\begin{figure}[!htb]
\begin{center}
{\includegraphics[width=9cm,clip=true, trim=0in 0in 0in 0.5in]{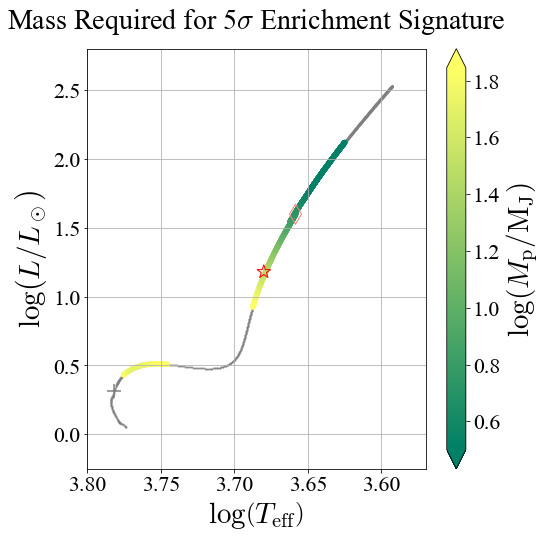}}
\caption{
The minimum-mass companion required to produce an ingestion-derived lithium enrichment signature with a $5{\sigma}$ statistical significance along the evolutionary track of KIC~\thekic{}.
At its current evolutionary phase, depicted as the red star, the ingestion of a 33~\mj{} brown dwarf companion is required to produce a statistically-significant lithium enrichment signature. 
The locations of the main sequence turnoff (grey cross) and luminosity bump (salmon diamond) are also illustrated.}
\label{Fig:min_MJ}
\end{center}
\end{figure}

To produce a statistically-significant ($5{\sigma}$) lithium-enrichment signature, a star of this mass and evolutionary state would be required to consume a companion of 33~\mj{}---a massive brown dwarf or tens of gas giants.
We illustrate the minimum-mass companion required to produce an ingestion-derived lithium enrichment signature with a $5{\sigma}$ statistical significance in Fig.~\ref{Fig:min_MJ}, where KIC~\thekic{} is illustrated in the H-R diagram as a red star. Also depicted is the location of the main sequence turnoff (grey-hued cross) and luminosity bump (salmon-hued diamond) corresponding to this stellar track.
Given the large mass that would need to be ingested to create a significant lithium enhancement, we determine that an observable lithium enrichment signature is extremely unlikely to be present in a star of this mass and evolutionary state, and therefore the lack of a lithium detection in this star is not sufficient evidence to rule out a significant interaction event. Similarly, the upper limit we are able to place on the lithium abundance is not low enough to argue for mixing driven lithium depletion \citep{Deal2015} and so the observed lithium constraint argues neither for nor against any particular formation scenario.

\section{Theoretical Explanations}
Given that our further investigations of the rotation of KIC \thekic\ seem to indicate that the inferred non-monotonic rotation profile is robust, 
it is interesting to consider how the star ended up in this unusual configuration.
There are several potential explanations that must be considered in our endeavour to understand the unusual rotation profile of KIC \thekic.

\subsection{Tidal Waves} \label{ssec:tides}
One exciting possibility would be the ingestion or interaction with a substellar body, perhaps an exoplanet. 
The \kepler\ mission in particular has emphasized that a large fraction of stars are orbited by at least one planet that is close enough in that it should be engulfed as the star expands on the red giant branch. Such an event may impact the surface rotation {\citep[e.g.][]{BolmontMathis2016,Privitera2016c,Qureshi2018,Stephan2020,Ahuiretal2021}}, and the lithium abundance \citep{AguileraGomez2016} of some stars, but is unlikely to cause other changes to the abundances in the surface convection zone. These predictions are consistent with our observations of KIC \thekic. If this is indeed a case of planetary ingestion, the existence of a rapidly rotating surface layer would seem to suggest that the angular momentum gain from the tidal interaction is happening from the outside in {layer by layer}, rather than treating the whole convection zone as a single {piece} \citep{Stephan2020}. Our detection of a star in this presumably transient non-monotonic state would also put significant constraints on the timescales and efficiencies of internal angular momentum transport mechanisms acting in this phase. 

{This is precisely what is expected if progressive (travelling) tidal waves are excited by a spiraling companion \citep{GoldreichNicholson1989}. These waves deposit their angular momentum at so-called critical layers, which correspond to co-rotation resonances between the local rotation frequency of the stellar fluid and the orbital frequency in the simplest case of a coplanar and circular orbit. This leads to a synchronisation of the interior of the star from its surface to the core. In the case of the convective envelope of KIC 9267654, the best candidate to explain such a phenomenon are tidal inertial waves \citep{OgilvieLin2007}, whose restoring force is the Coriolis acceleration. Therefore, when they meet such a co-rotation layer, they are able to transfer the momentum they carry to the stellar (differential) rotation \citep{Favieretal2014,Astouletal2021}. This would have significant implications for the estimation of tidal dissipation quality factors and the mechanisms of angular momentum transport in the convective zones of stars hosting companions \citep{AstoulBarker2021}. 

Since planetary engulfment and associated tidal flows are potentially the best candidates to explain the observed subsurface acceleration, each category of tidal flows should be examined. First, the gravitational perturbation by a potential companion triggers large-scale hydrostatic/equilibrium flows \citep{Zahn1966,Terquemetal1998,remus2012,Ogilvie2013}. Considering the mathematical expressions derived in \cite{remus2012} and \cite{Terquem2021}, one can demonstrate that the latitudinally-averaged flux of angular momentum carried by Reynolds stresses along the vertical direction of these equilibrium flows vanishes. Therefore, these flows cannot explain the needed transport of angular momentum and one should consider the complementary tidal flows: the dynamical tide. The dynamical tide is constituted of waves that propagate in stars (and planets) that can be excited by  tidal interactions with a companion \citep[][]{Zahn1975,OgilvieLin2004}. In a stellar convective envelope, the dynamical tide is constituted by tidal inertial waves excited by the Coriolis acceleration of the equilibrium tide \citep{OgilvieLin2007,Ogilvie2013}. As has been pointed out at the beginning of this discussion, they are able to transport angular momentum thanks to their Reynolds stresses and critical layers, leading to a synchronisation (i.e. an acceleration) of the convective envelope from the surface to the interior. For a companion spiralling toward the star when the orbital mean motion $n_{\rm orb}$ is greater than the stellar surface rotation frequency, $\Omega_{\rm s}$, 
these waves are excited when $P_{\rm orb}>P_{\rm s}/2$ (e.g. \citealt{BolmontMathis2016} with $P_{\rm orb}=2\pi/n_{\rm orb}$ the orbital period and $P_{\rm s}=2\pi/\Omega_{\rm s}$ the rotation period of the surface of the star). Using scaling laws for the amplitude of the equilibrium tide that scales as $1/a^3$ \citep{remus2012,Ogilvie2013}, where $a$ is the semi-major axis, one can demonstrate that {the} Reynolds stresses {of tidal inertial waves, which are excited by the Coriolis acceleration of the equilibrium tide \citep{Ogilvie2013},} scale as $1/a^6$. {They thus strongly increase} with the approach of the companion. One challenging remaining question is {to quantify} the non-linear interactions between the flows of the turbulent convection, of the hydrostatic equilibrium tide, and of the dynamical tides \citep[e.g.][]{Duguidetal2021,Terquem2021,BarkerAstoul2021} and the resulting angular momentum transport \citep{Astouletal2021}}, another is to compare the normal oscillations we see in this system to the close binary stars with suppressed oscillation amplitudes from \citet{Gaulme2014}.  We do not attempt to study the details of such an interaction here, but suggest that this system might represent an interesting test case for future {theoretical models and numerical simulations}.

\subsection{Turbulent Convection}
To explain the observed subsurface acceleration, one can also examine the mechanisms that transport angular momentum in single stars. The first one is turbulent convection. Depending on the so-called convective Rossby number, different regimes of differential rotation can be triggered \citep{Brunetal2017}. {On the one hand, this dimensionless number can be defined as the ratio of the rotation period to the convective overturn time in stellar evolution theory. This is the stellar convective Rossby number $Ro_{\rm s}$ \citep[e.g.][]{landin2010}. On the other hand, it can be defined as the ratio between the vorticity of convective flows and twice the rotation angular frequency of the star in (magneto-)hydrodynamical numerical simulations of stellar convection zones. This is the so-called fluid convective Rossby number $Ro_{\rm f}$ \citep[][]{Brunetal2017}. These Rossby numbers compare the inertia of convective flows to their Coriolis acceleration; their relationships have been discussed in \cite{Brunetal2017}, who showed that they are of the same order of magnitude (we refer the reader to Table 4 and Appendix B in that article) and who identified the different regimes of differential rotation as a function of $Ro_{\rm f}$}. First, for small convective Rossby numbers (fast rotation; $Ro_{\rm f}\le 0.3$), the Coriolis acceleration constrains convective flows through the so-called Taylor-Proudman constraint \citep[e.g.][]{Rieutord}. This leads to cylindrical differential rotation profiles like those observed at the surface of Jupiter and Saturn. Next, for moderate convective fluid Rossby numbers of the order of unity, conical differential rotation profiles are established with a solar-like equatorial acceleration if $0.3<Ro_{\rm f}\le 1$ and an anti-solar polar acceleration if $Ro_{\rm f}\ge 1$. Finally, when $Ro_{\rm f}$ becomes large, the action of the Coriolis acceleration becomes negligible and shellular-like differential rotation profiles depending mainly on the radius can be obtained \citep{BrunPalacios2009}. {Note that this classification has been established in an hydrodynamical framework and that \cite{Brunetal2022} showed how it can be modified in presence of a dynamo-generated magnetic field. In particular, the feed-back of the Lorentz force weakens the absolute value of the differential rotation, in particular in the case of fast-rotating stars. In addition, the transition between the conical solar-like differential rotation and the anti-solar differential rotation state is particularly sensitive to the complex properties of turbulent convective flows \citep{Gastineetal2014,Kapylaetal2014}.} We compute the value of $Ro_{\rm s}$ in KIC 9267654 and we obtain that $Ro_{\rm s}>7$ in the bulk of the envelope while $Ro_{\rm s}\sim 1$ at the surface. Therefore, KIC 9267654 seems to be in the latter regime. In this context, the 3D nonlinear global spherical simulations of the dynamics of the convective envelope of a red giant star computed by \cite{BrunPalacios2009} with the ASH (Anelastic Spherical Harmonics) code \citep{Cluneetal1999,Brunetal2004} is interesting. Indeed, it shows how a shellular differential rotation can be established when $Ro_{\rm f}$ is large. In that case, the obtained latitudinally-averaged differential rotation radial profile is decreasing towards the surface (we refer the reader to the right column of Fig. 10 in \citealt{BrunPalacios2009}), a trend that is conserved in \cite{BrunPalacios2009} when increasing the rotation rate that leads to anti-solar differential rotation profiles. We are not aware of evidence that the transition in convective Rossby number close to the surface can lead to any observed acceleration of the rotation rate consistent with what we observe for KIC \thekic. Therefore, we consider it more likely that the rotation profile is explained by a planetary engulfment event.

\subsection{Internal Gravity Waves}
The last mechanism that could be considered in single stars are travelling (progressive) internal gravity waves and mixed (gravito-acoustic) modes that are efficiently stochastically excited by turbulent convection in evolved low-mass stars \citep[e.g.][]{TalonCharbonnel2008,Pinconetal2016,Pinconetal2017,Dupret2009}. Indeed \cite{Schatzman1993}, \cite{Zahnetal1997}, \cite{Talon2005}, \cite{Rogers2015} and \cite{Belkacemetal2015a,Belkacemetal2015b} have demonstrated that they are able to efficiently transport angular momentum. However, this statement is only true for stably stratified stellar radiation zones where 
internal gravity waves {propagate} and gravity modes {form}; {they} are evanescent in stellar convective regions. In addition, acoustic modes are not efficient at redistributing angular momentum in convective regions. As a consequence, these mechanisms should be excluded from further discussion of KIC \thekic\, where the unexpected differential rotation is localized within the surface convection zone.\\

\section{Discussion}
We have presented evidence to support the existence of a nonmonotonic radial rotation profile for KIC \thekic, with a rapidly rotating core (P$_\textrm{core}=12.6$ days), a slowly rotating envelope (P$_\textrm{env} \gtrsim 178$ days), and a moderately rotating surface (P$_\textrm{surf}=42$ days), inferred using a combination of spectroscopic and asteroseismic data. We have collected this evidence from two entirely independent spectroscopic investigations, as well as several asteroseismic analyses. The data seem to consistently indicate that KIC \thekic\ is a relatively normal lower red giant branch star, whose only glaringly unusual feature is its odd rotation profile, and possibly a few offset abundances. While \gaia\ suggests a companion in an orbit much longer than the rotation period, 
we find no evidence for the existence of a tidally synchronized stellar companion or significant ongoing mass transfer in this system, although we cannot exclude such scenarios entirely.

We note that this star is not the only giant with an unusual surface rotation rate. Other stars have been noted whose claimed surface rotation rates are faster than their core rotation rate \citep[e.g.][]{Kurtz2014,Ceillier2017, Tayar2019b, Aerts2019}, or significantly faster than single stellar evolution would predict \citep[e.g.][]{Ceillier2017, TayarPinsonneault2018}. However, in most of these cases, the unusual surface rotation is presumed to be linked to tidal interactions with another star \citep{Tayar2015,LiG2020, MazzolaDaher2021} or a much earlier planetary engulfment \citep{Carlberg2012,BolmontMathis2016,Privitera2016c,Ahuiretal2021}. In other cases, these stars come from ensemble analyses, where a few percent of stars are often affected by measurement errors or target contamination \citep{Aigrain2015}, and envelope rotation rates are rarely available. Finally, in other well {studied red giants including e.g. KIC 3744043, the envelope rotation rate estimated with asteroseismic techniques (P$_{\rm env} \sim$ 104 - 159 days; \citealt{Triana2017}) may come out slightly slower than the surface rotation rate measured from broadening of spectroscopic lines (\vsini=2.5$\pm$1.0 \kms, P$_{\rm rot} \sim  94.5 \pm 45.9$ days; \citealt{Thygesen2012, Gehan2021}).
Nonetheless, the disagreement is within the properly propagated measurement uncertainties and its core is rotating significantly faster than both its envelope and surface, at a period of P$_{\rm core} \sim$ 21.6 days \citep{Triana2017,Gehan2018}, which makes it hard to draw strong conclusions.}  
What makes KIC \thekic\ so interesting is the existence of a well-measured three-point rotation profile where all three measurements have been verified by multiple teams and independent analyses, and the surface and envelope rotation rates disagree with each other at a significant level given the small measurement uncertainties.

Given the exciting possible explanations of the unusual rotation profile detected in KIC \thekic, and their impact on our understanding either of planetary engulfment and tidal theory or some unexplained physics in low-mass stellar interiors, we encourage further observational and theoretical work on this exciting system. In particular, more detailed observational analysis of the surface rotational and chemical properties including r-process and s-process abundances could be interesting, as would longer term searches for spot modulation as well as searches for surviving planetary companions in the system. Theoretical modeling of both tidal theory in general, as well as specific companions and orbital configurations that could produce the characteristics observed for KIC \thekic\ might prove quite illuminating. More generally, the ability to not only identify planets in stable configurations similar to our own solar system, but also to trace out their changes as their stars evolve \citep{Rizzuto2017,Grunblatt2017, Chontos2019,Gansicke2019} has opened and is likely to continue to open interesting windows toward understanding planetary formation, evolution, and long-term habitability, while simultaneously constraining the physics of stellar surfaces and interiors.

\begin{acknowledgements}

We thank Carles Badenes, Christine Daher, Daniel Huber, Anne Hedlund, and Lisa Bugnet for helpful discussions that contributed to this manuscript. We also thank the members of the rotation working group at Tthe KITP Transport in Stars 2021 program for input on this paper, and therefore acknowledge that this research was supported in part by the National Science Foundation under Grant No. NSF PHY-1748958.

Support for this work was provided by NASA through the NASA Hubble Fellowship grant No.51424 awarded by the Space Telescope Science Institute, which is operated by the Association of Universities for Research in Astronomy, Inc., for NASA, under contract NAS5-26555. S.K. acknowledges funding from the European Union H2020-MSCA-ITN-2019 under grant agreement no. 860470 (CHAMELEON).

DMB and TVR gratefully acknowledge funding from the Research Foundation Flanders (FWO) by means of senior and junior postdoctoral fellowships with grant agreements 1286521N and 12ZB620N, respectively, and FWO long stay travel grants V411621N and V414021N, respectively.

MSF gratefully acknowledges support provided by NASA through Hubble Fellowship grant HST-HF2-51493.001-A awarded by the Space Telescope Science Institute, which is operated by the Association of Universities for Research in Astronomy, In., for NASA, under the contract NAS 5-26555. 

R.A.G., S.N.B and S.M.\ acknowledge the support from GOLF and PLATO CNES grants at CEA Paris-Saclay.

SLM acknowledges the support of the Australian Research Council through Discovery Project grant DP180101791.

S.R. gratefully acknowledges support of the DFG priority
program SPP 1992 “Exploring the Diversity of Extrasolar Planets” (RE~2694/7-1).

Based on observations obtained with the HERMES spectrograph, which is supported by the Research Foundation – Flanders (FWO), Belgium, the Research Council of KU Leuven, Belgium, the Fonds National de la Recherche Scientifique (F.R.S.-FNRS), Belgium, the Royal Observatory of Belgium, the Observatoire de Genève, Switzerland and the Thüringer Landessternwarte Tautenburg, Germany.

This paper includes data collected by the Kepler mission and obtained from the MAST data archive at the Space Telescope Science Institute (STScI). Funding for the Kepler mission is provided by the NASA Science Mission Directorate. STScI is operated by the Association of Universities for Research in Astronomy, Inc., under NASA contract NAS 5–26555.

This work has made use of data from the European Space Agency (ESA) mission
{\it Gaia} (\url{https://www.cosmos.esa.int/gaia}), processed by the {\it Gaia}
Data Processing and Analysis Consortium (DPAC,
\url{https://www.cosmos.esa.int/web/gaia/dpac/consortium}). Funding for the DPAC
has been provided by national institutions, in particular the institutions
participating in the {\it Gaia} Multilateral Agreement.

This research has made use of NASA's Astrophysics Data System Bibliographic Services.

Funding for the Sloan Digital Sky 
Survey IV has been provided by the 
Alfred P. Sloan Foundation, the U.S. 
Department of Energy Office of 
Science, and the Participating 
Institutions. 

SDSS-IV acknowledges support and 
resources from the Center for High 
Performance Computing  at the 
University of Utah. The SDSS 
website is www.sdss.org.

SDSS-IV is managed by the 
Astrophysical Research Consortium 
for the Participating Institutions 
of the SDSS Collaboration including 
the Brazilian Participation Group, 
the Carnegie Institution for Science, 
Carnegie Mellon University, Center for 
Astrophysics | Harvard \& 
Smithsonian, the Chilean Participation 
Group, the French Participation Group, 
Instituto de Astrof\'isica de 
Canarias, The Johns Hopkins 
University, Kavli Institute for the 
Physics and Mathematics of the 
Universe (IPMU) / University of 
Tokyo, the Korean Participation Group, 
Lawrence Berkeley National Laboratory, 
Leibniz Institut f\"ur Astrophysik 
Potsdam (AIP),  Max-Planck-Institut 
f\"ur Astronomie (MPIA Heidelberg), 
Max-Planck-Institut f\"ur 
Astrophysik (MPA Garching), 
Max-Planck-Institut f\"ur 
Extraterrestrische Physik (MPE), 
National Astronomical Observatories of 
China, New Mexico State University, 
New York University, University of 
Notre Dame, Observat\'ario 
Nacional / MCTI, The Ohio State 
University, Pennsylvania State 
University, Shanghai 
Astronomical Observatory, United 
Kingdom Participation Group, 
Universidad Nacional Aut\'onoma 
de M\'exico, University of Arizona, 
University of Colorado Boulder, 
University of Oxford, University of 
Portsmouth, University of Utah, 
University of Virginia, University 
of Washington, University of 
Wisconsin, Vanderbilt University, 
and Yale University.

\end{acknowledgements}

\software{\texttt{apollinaire} \citep{Breton2022}, Astropy \citep{Astropy,Astropy_2018}, corner \citep{corner}, george \citep{george_python}, h5py \citep{hdf5_python}, Matplotlib \citep{Matplotlib}, NumPy \citep{numpy}, pandas \citep{reback2020pandas}, SciPy \citep{2020SciPy-NMeth}}

\facilities{Du Pont (APOGEE), Sloan (APOGEE), 2MASS,  \emph{Gaia}, \emph{Kepler}, HERMES}

\bibliographystyle{aasjournal}
\bibliography{ms, library, library2}

\end{document}